\documentclass[a4paper]{emulateapj}
\usepackage{apjfonts}
\usepackage{epsfig,amsmath,natbib}
\usepackage[dvips]{color}
\usepackage{verbatim}

\newcommand{\fig}[1]{Fig.\ \ref{#1}}

\newcommand{\keV}{\textrm{keV}}

\newcommand{\cm}{\textrm{cm}}
\newcommand{\rhe}{r_{180}}

\definecolor{Black}{named}{Black}
\definecolor{Red}{named}{Red}

\begin{document}

\shorttitle{SZ profiles for clusters of galaxies}
\shortauthors{T. Haugb{\o}lle, J. Sommer-Larsen, K. Pedersen}

\title{Sunyaev-Zeldovich profiles for clusters and groups of galaxies}

\author{Troels Haugb\o lle\altaffilmark{1,2}, Jesper Sommer-Larsen\altaffilmark{3,1},
Kristian Pedersen\altaffilmark{1}}
\altaffiltext{1}{Dark Cosmology Centre, Niels Bohr Institute, University of Copenhagen,
Juliane Maries Vej 30, DK-2100 Copenhagen, Denmark}
\altaffiltext{2}{Department of Physics and Astronomy, University of Aarhus, DK-8000
Aarhus C, Denmark}
\altaffiltext{3}{Excellence Cluster Universe, Technische Universit\"at
       M\"unchen, Boltzmanstr. 2, D-85748 Garching, Germany}
\email{troels_h@dark-cosmology.dk}
\email{jslarsen@astro.ku.dk}
\email{kp@dark-cosmology.dk}

\begin{abstract}
The Sunyaev-Zeldovich (SZ) effect gives a measure of the thermal energy and electron pressure in groups
and clusters of galaxies. In the near future SZ surveys will map hundreds of systems, shedding light on
the pressure distribution in the systems. The thermal energy is related to the total mass of a system of
galaxies, but it is only a projection that is observed through the SZ effect. A model for the 3D distribution
of pressure is needed to link the SZ signal to the total mass of the system.
In this work we construct an empirical model for the 2D and 3D SZ profile, and compare it to a set
of realistic high resolution SPH simulations of galaxy clusters and groups, and to a stacked SZ
profile for massive clusters derived from WMAP data. Furthermore, we combine observed
temperature profiles with dark
matter potentials to yield an additional constraint, under the assumption of hydrostatic equilibrium.
We find a very tight correlation between the characteristic scale in the model, the integrated SZ
signal, and the total mass in the systems with a scatter of only 4\%. The model only contains two free
parameters, making it readily applicable even to low resolution SZ observations of galaxy clusters.
A fitting routine for the model that can be applied to observed or simulated data
can be found at \url{http://www.phys.au.dk/~haugboel/software.shtml}.
\end{abstract}

\subjectheadings{}

\section{Introduction}
When Cosmic Microwave Background (CMB) photons pass through galaxy clusters, they Compton
up-scatter on hot electrons, making a small increment (decrement) above (below)
the peak of the CMB primary spectrum. The size of this distortion in the CMB spectrum, the
Sunyaev-Zeldovich (SZ) effect
\citep{SZ:1972}, is proportional to the electron pressure integrated along the line-of-sight.
Besides being an important probe of the physics of the intracluster medium, the SZ effect is a
promising tool in cosmology: The signal produced by a cluster is practically redshift
independent and can be observed to high redshifts. The Planck surveyor satellite \citep{planck}
will produce a cluster catalogue with up to $\sim$10.000
(sufficiently massive) clusters out to $z\sim 1$ \citep{schaefer:2006}.
By combining SZ and X-ray observations of relaxed clusters the Hubble constant $H_0$ can
be derived, limits can be put on $\Omega_m$, and the gas mass fraction in
clusters can be measured \citep{Bonamente:2005,LaRoque:2006}.
The SZ effect was first detected in three nearby clusters more than two decades ago by \cite{Birkishaw1984}, but
the last five years, due to advances in sub millimetre receiver technology, routine measurements
have been done for $\mathcal{O}(100)$ clusters using facilities such as the OVRO and BIMA
telescopes. Right now a large number of telescopes have started observing the SZ signal or are under
construction
(e.g.~ACBAR, CARMA, SUZIE III, SPT, APEX etc.), and in the near future ALMA
will gradually become online enabling unprecedented resolution and sensitivity for making detailed
observations of individual clusters. Current SZ surveys of galaxy clusters 
\citep[see e.g.][for a set of current observations]{Bonamente:2006} have observed
the unresolved integrated SZ signal from clusters, but with CARMA, ALMA and the SPT also
the SZ signal as a function of radius from the cluster centre will be measured. \cite{Afshordi:2006} has
used the 3rd year WMAP data to extract SZ images from 193 clusters. Stacking them they have obtained
an averaged SZ profile for clusters with $T_X>3 $keV.

Spatially resolved SZ images probe, in a way complementary to X-ray
observations, the distribution of thermal energy in clusters which in turn
is closely linked to the total cluster mass. Furthermore, since the SZ effect
essentially depends on the electron pressure, the physics going into
understanding it is simpler and more robust, than is the case for the X-ray emission.

In this paper we use realistic high resolution N-body/SPH simulations of galaxy clusters and groups
together with observed temperature profiles of nearby groups and clusters to predict the radial SZ
profile of different types of systems of galaxies, and compare our results to the
averaged profile of \cite{Afshordi:2006}. We introduce a universal fitting formula, with only two free
parameters, that can be employed in future observations. Furthermore, we show that fitting this
profile enables a precise estimate of the total mass in the system.
In the next section we describe the computer experiments, that are used to construct the synthetic
SZ profiles. In section 3 we discuss the simulated data and present our results, and in section 4
we discuss and provide our conclusions.

\section{Computer experiments}
We use 24 galaxy group and galaxy cluster simulations to study the SZ effect
for systems of virial temperatures of about 1 to 6 keV. The models include 12 groups 
of approximately the same mass ($M_{vir} = 0.8-1.1 \times 10^{14} M_\odot$), 
11 identical clusters with $M_{vir} = 2.7\times 10^{14} M_\odot$ (``Virgo'' clusters), 
but simulated with different gas physics, and a single large cluster of
$M_{vir} = 12\times 10^{14} M_\odot$ (a ``Coma'' cluster).
The models are re-simulated from a low resolution cosmological dark matter simulation, where
the halos are identified with a halo finder. The particles are traced back in time to a initial redshift
$z_{initial}$, and the virial volume is repopulated with both gas and dark matter at a high resolution.
The code includes radiative cooling, star formation, supernova feedback, 
chemical evolution and back reaction from a redshift dependent UV field. The different models 
are summarised in Table \ref{tab:models}. For a detailed description of the different gas physics
going into the code in general, and the ``Virgo'' cluster simulations
in particular we refer to \citet{Romeo:2006}. The groups were all selected at 
random, the only criterion being their virial mass, and therefore they comprise a cosmological
fair sample of groups with $M_{vir}=0.8-1.1 \times 10^{14} M_\odot$.

The different masses of the clusters allow us to probe the mass dependence of the SZ profile,
while the  ``Virgo'' models are used to investigate how robust our predictions are for the SZ profile
with respect to assumptions on the underlying physics. The groups, being a statistically unbiased sample,
give limits on cosmic variance for the given virial mass ($10^{14} M_\odot$). 

We have divided the groups into three different classes according to their morphology and evolutionary 
history: Groups with merging activity (186,231,239,262), fossil groups (189,228,236,244) and normal
groups (190,233,247,276).
A group is considered fossil if the difference in apparent R-band magnitude of the first and second
brightest galaxy is greater than two \citep{Jones:2000}, and a group is merging if it by visual inspection
has significant merging activity in the core, or if the rms scatter in 3D radial shells of the temperature, pressure and density, is comparable to the average value in the shell.
\begin{table}[h] \begin{center} \begin{tabular}{@{}c@{}c@{}c@{}c@{}c@{}}
\hline \hline Name&$T_{X} [\keV]\,\,$&$\,\,T_{mg} [\keV]\,\,$
               &$\,\,M_{180} [10^{14} M_\odot]$&Comments\cr
\hline Groups & \multicolumn{4}{c}{Re-simulated groups with $z_{initial}=39$}\cr \hline 
186		& 0.99 & 1.08 & $0.92$ & Merging group\cr
189		& 1.07 & 1.09 & $0.91$ & Cool-Core, Fossil group\cr
190		& 1.21 & 1.22 & $0.99$ & Normal group \cr
228		& 1.07 & 1.10 & $0.85$ & Cool-Core, Fossil group\cr
231		& 0.99 & 1.08 & $0.84$ & Merging group\cr
233		& 1.03 & 1.09 & $0.82$ & Normal group \cr
236		& 1.00 & 0.99 & $0.73$ & Fossil group\cr
239		& 0.91 & 1.11 & $0.81$ & Merging group\cr
244		& 1.13 & 1.08 & $0.80$ & Fossil group\cr
247		& 1.01 & 1.15 & $0.90$ & Normal group\cr
262		& 0.97 & 1.03 & $0.88$ & Merging group\cr
276		& 1.02 & 1.07 & $0.82$ & Cool-Core group\cr \hline
Clusters & \multicolumn{4}{c}{Re-simulated clusters with $z_{initial}=19$}\cr \hline
AY-SW	& 2.13 & 2.07 & $2.32$ & Arimoto-Yoshii IMF\cr	
AY-SW-8	& 2.18 & 2.10 & $2.36$ & 8 times resolution\cr		
AY-Vol39& 2.18 & 2.18 & $2.36$ & $z_{initial} = 39$ \cr		
AY-SWx2 & 2.18 & 2.04 & $2.40$ & 2 times SNII feedback \cr	
AY-SWx4 & 2.17 & 2.09 & $2.38$ & 4 times SNII feedback \cr	
AY-PH0.75&2.23 & 2.10 & $2.41$& preh. $0.75 \keV/$part @ z=3 \cr	
AY-PH1.5 &2.27 & 2.14 & $2.37$ & preh. $1.5 \keV/$part @ z=3 \cr 
AY-PH50	& 2.19 & 2.07 & $2.41$ & preh. $50 \keV \cdot \cm^2$part @ z=3 \cr	
AY-COND & 2.26 & 2.08 & $2.43$ & Thermal conduction \cr	
Sal-SW	& 2.24 & 2.10 & $2.42$ & Salpeter IMF \cr			
Sal-WFB	& 2.27 & 2.23 & $2.41$ & Weak feedback \cr		
Coma	& 5.57 & 5.27 & $10.4$ &Relaxed massive cluster \cr	
\hline \hline
\end{tabular} \end{center}
\caption{Characteristics of simulated clusters} \label{tab:models}
\end{table}

\begin{figure*}
\begin{center}
\includegraphics[width=0.95 \textwidth]{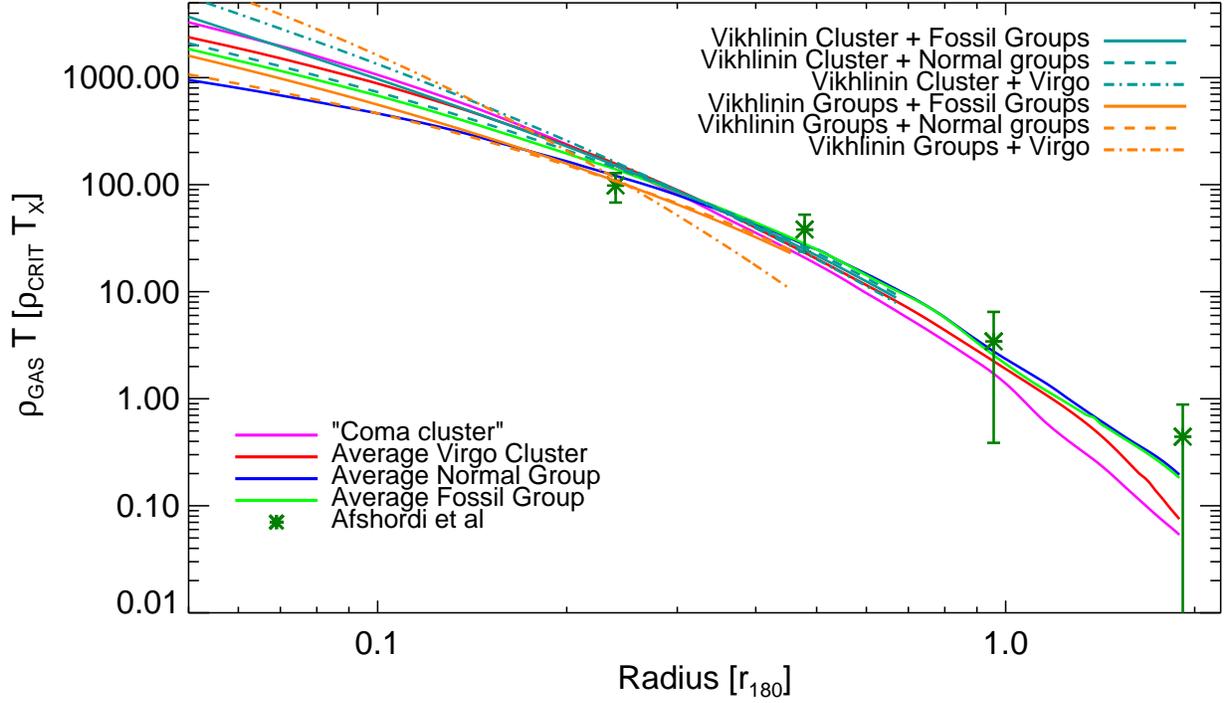}
\caption{Spherically averaged 3D SZ profiles computed from the simulations, the
stacked profile of \cite{Afshordi:2006}, and derived
3D profiles from observed X-ray temperatures combined with average
gravitational potentials. The profiles are normalised to the critical density $\rho_{CRIT}$,
and the typical temperature of the system, $T_X$, defined as the average projected spectral-like
temperature in the interval $0.1\,r_{200} < r < 0.4\,\,r_{200}$.
To reduce the visual scatter only average profiles from the simulations
are shown.}\label{fig:szeprofs}
\end{center}
\end{figure*}

\section{Analysis}
The SZ profiles of the simulated systems can only be used as templates for an universal fitting
formula, if the profiles are in accordance with observations. In \fig{fig:szeprofs} are shown
the spherically averaged SZ profiles compared to the only currently published
SZ profile (\citet{Afshordi:2006}), obtained using data from the WMAP satellite.
The observed points may indicate a slightly more bended profile than the simulations,
but within the 1-$\sigma$ error bars there is fairly good agreement. From the simulations there is a clear
trend towards steeper profiles, for more massive and relaxed clusters and groups reflecting differences in
the underlying DM potentials (see \fig{fig:dmpot}). This should be recalled when constructing an
average observational profile, because the average profile may not represent a true (``universal'')
physical profile, but rather a smeared average of the real profiles.

\subsection{The inner part of the Sunyaev-Zeldovich profile}
The central core of the SZ profile ($r<0.25\,\rhe$ cannot be probed by WMAP, due to its limited resolution
of at most $0.12^\circ$. To extend the dynamic range of the observations we combine gravitational
potentials from the simulations with observed average X-ray temperature profiles. Under the assumption
of hydrostatic equilibrium in the core of the systems, we can then predict the inner part of the SZ profile.

\begin{figure}
\begin{center}
\includegraphics[width=0.45 \textwidth]{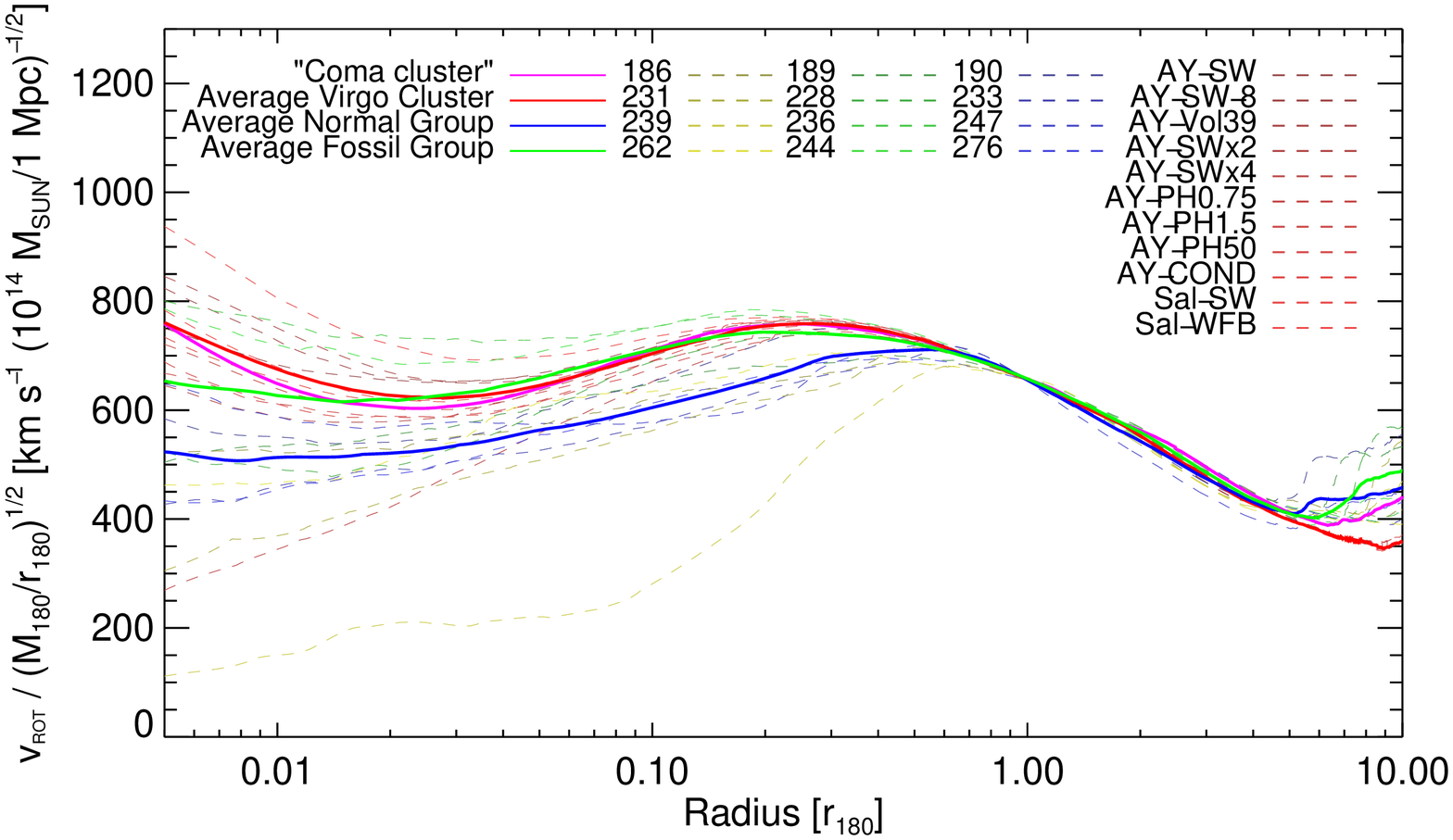}
\includegraphics[width=0.45 \textwidth]{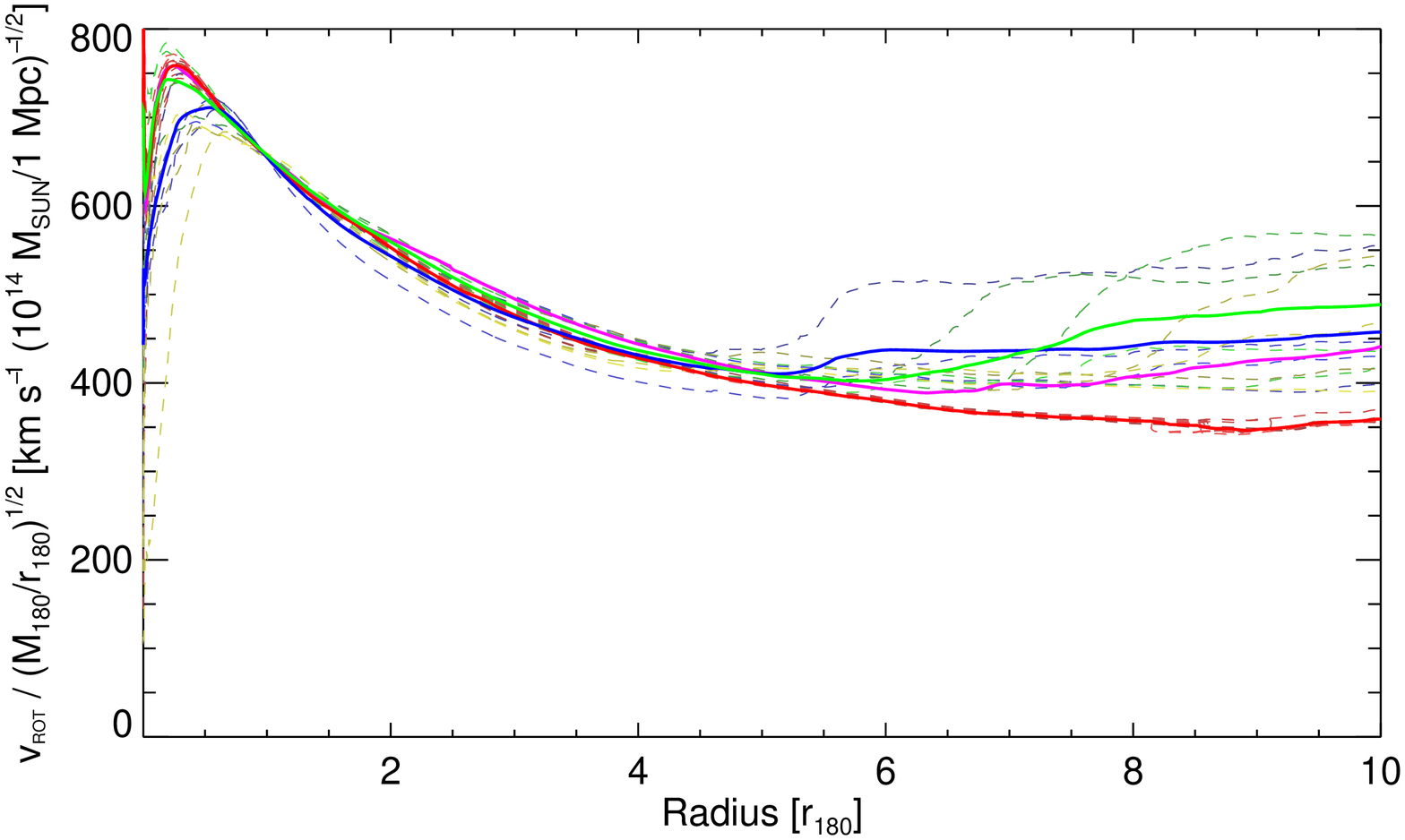}
\caption{Dark matter potentials measured as the equivalent rotational speed, both for specific
groups and clusters, and for the average ``Virgo'' cluster, normal and fossil group. The curves are
normalised with the virial mass.}\label{fig:dmpot}
\end{center}
\end{figure}

The Compton $y$-parameter, which determines the overall temperature decrement in the CMB radiation
due to the SZ effect, is proportional to the integrated pressure $P$ of the electrons along the line of sight
\begin{equation}
y = \int y(l) dl = \int \frac{\sigma_T P(l)}{m_e c^2} dl = \int \sigma_T n_e \frac{k_B T_e}{m_e c^2} dl\,.
\end{equation}
Assuming hydrostatic equilibrium and spherical symmetry
\begin{equation}
\frac{dP}{d r} = -\rho_g \frac{GM(r)}{r^2}
\end{equation}
we can use the proportionality of $P$, and the Compton $y$-parameter in a volume element,
$y(l(r))$, to obtain
\begin{equation}\label{eq:hs}
\ln \frac{y(r)}{y(r_0)} = \ln \frac{P(r)}{P(r_0)} = - \mu_g m_p \int^r_{r_0} \frac{GM(r')}{k_b T r'^2} dr' 
\end{equation}
We have constructed three different averaged gravitational potentials based on the normal groups, the
fossil groups and the ``Virgo'' clusters. For the temperature profiles we use observed clusters by \citet{Vikhlinin:2005}, and make one average profile constructed from
the galaxy groups with
virial temperatures less than 2.5 keV, and one constructed from the massive clusters in the sample with
virial temperatures between 3.5 and 8.5 keV. 
The dark matter distribution, derived from the simulations, is quite robust against the specific gas
physics involved in the simulation. This is demonstrated in \fig{fig:dmpot}, where it is seen
that all the ``Virgo'' models, in spite of the different gas physics, have essentially
the same gravitational potentials from $0.02\,\rhe$ and outwards. However, there are systematic
differences between different types of systems with the same mass. Fossil groups are more
relaxed compared to normal groups, and their mass distribution is similar to relaxed clusters, like the
``Virgo'' and ``Coma'' models from $0.02\,\rhe$ and outwards (see \fig{fig:dmpot}). Therefore
the above reconstruction procedure gives a good idea of future resolved SZ
observations of the cores of clusters of galaxies, and our six resolved profiles are a fair sample of
what can be expected.

\begin{figure}
\begin{center}
\includegraphics[width=0.45 \textwidth]{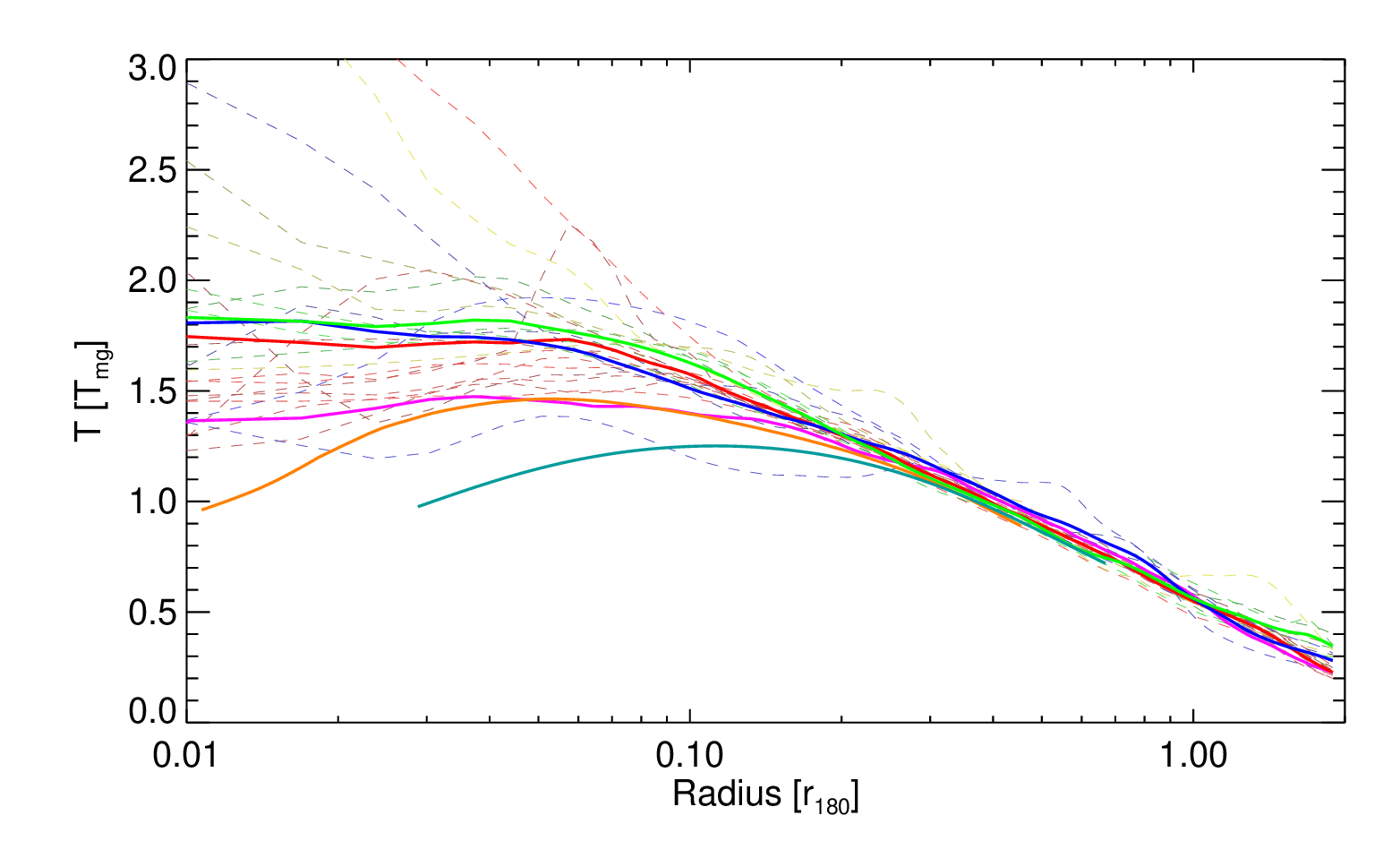}
\includegraphics[width=0.45 \textwidth]{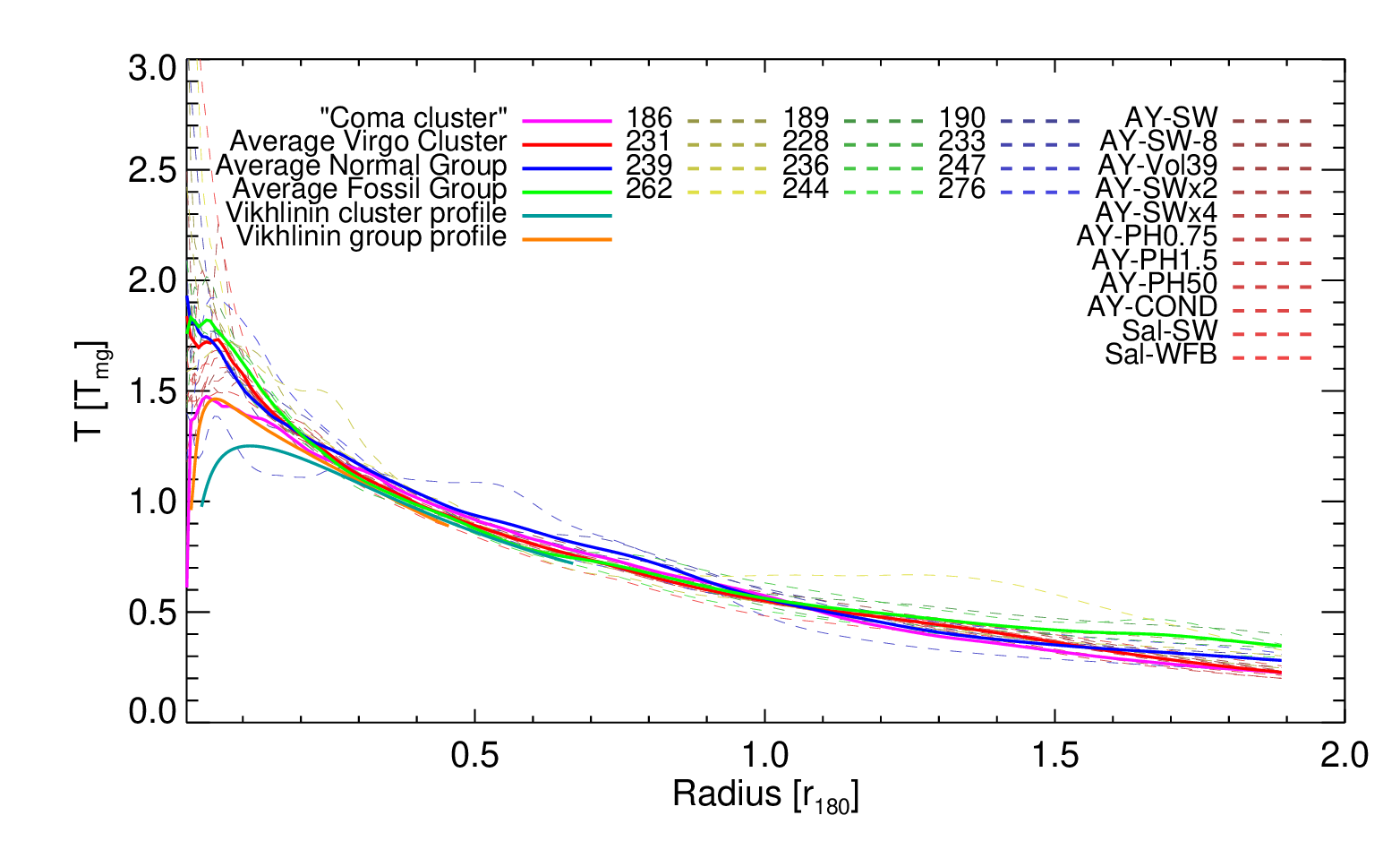}
\caption{Mass-weigthed 3D temperature profiles for the simulated systems
compared to observed an average group profile (1.6 keV$<T_X<$ 2.6 keV), and an
average cluster profile (3.5 keV $<T_X<$ 8.5 keV).}\label{fig:tempprof}
\end{center}
\end{figure}

\subsection{Temperature profiles}
From \fig{fig:szeprofs} it is clear, that albeit the SZ profiles
constructed from the observed temperature profiles and the dark matter
density profiles are in approximate agreement with the profiles extracted
from the simulations, they have a tendency to be more peaked at the
centre. This can be traced to differences in the observed and simulated temperature
profiles (see e.g.~Eq.~\ref{eq:hs}), and to the typical masses of the observed systems.
The different temperature profiles are normalised with a global temperature $T_X$. To 
enable direct comparison to the observed temperature profiles by
\cite{Vikhlinin:2005} we compute $T_X$ as the average projected spectral-like temperature
\citep{Rasia:2005} in the interval $0.1\,r_{200} < r < 0.4\,\,r_{200}$. All profiles converge to the
same universal curve for $r>0.2\,\rhe$, but in the inner part of the systems there
are differences (see \fig{fig:tempprof}). The simulated systems have all flat profiles at the
core, which are in good qualitative agreement with observations of clusters of similar virial mass,
but the average observed profiles have lower normalised temperatures towards the centre,
with more or less the same offset between the ``Coma'' cluster and the observed cluster profile
(constructed from clusters with 3.5 keV $< T_X <$ 8.5 keV),
and between the average simulated and observed group temperature profiles
(the latter constructed from groups with $T_X < 2.5$ keV \citet{Vikhlinin:2006}).
The offset compared to observations and relatively flat temperature profiles are well known
problems for simulations invoking radiative cooling and feedback processes
\citep[e.g.][]{Borgani:2004, Pratt:2006}, but it is further accentuated by an offset in mass
between the observed and simulated groups and the observed clusters, and the
simulated ``Virgo'' clusters.

\subsection{A universal SZ profile}
As can be seen in \fig{fig:szeprofs} all the SZ profiles have nearly the same form,
but the overall normalisation, tension and slope of the different profiles
depends on the mass, and the specific cluster. To construct a simple model we have tried to fit a
variety of combinations of beta profiles and exponential profiles 
for the density combined with a polytropic or isothermal equation of state, but it does not yield a
satisfactory fit. The correct temperature to use, when constructing a SZ profile, is the
mass weighted, and it does not necessarily agree with the temperature inferred from X-ray
observations (\cite{Bonaldi:2007,Afshordi:2007}), which may explain why the above combination
works well for X-ray observations, while not so for our sample of SZ profiles.

To circumvent this problem we are using a novel universal profile that directly fit the data, 
without assuming anything about the underlying temperature and density distributions, while
using as few parameters as possible. The main morphological differences are in the
slope/tension and in the normalisation of each profile, and it is indeed
possible to construct a model, that with only two free parameters, a normalisation, and a
characteristic scale can fit the full set of simulations and observations in detail.

The profiles extracted from the simulated data do not per se contain any
errors, but a measure of the natural scatter in a profile $y(r)$ at a given radial
distance, is the variance $\sigma(r)$ of $y(\bf r)$ inside the radial bin at $r$.
We have used this variance estimate to construct a goodness of fit parameter
$\chi^2$ for each projected 2D and spherically averaged 3D profile $y_{2D}(r)$,
$y_{3D}(r)$ given by
\begin{equation}
\chi^2_{2D,3D} = \frac{1}{r_{max} - r_{min}} \int_{r_{min}}^{r_{max}}
    \frac{\left(y_{2D,3D}(r) - y^m_{2D,3D}(r,\bf p)\right)^2}{\sigma_{2D,3D}^2(r)} dr\,,
\end{equation}
where $y^m_{2D,3D}(r,\bf p)$ is the model with parameters $\bf p$. The internal scatter
in relaxed systems is much smaller, giving stronger constraints than from merging systems.
This is sensible, since the observational scatter among relaxed clusters is smaller too.

\begin{figure}
\begin{center}
\includegraphics[width=0.45 \textwidth]{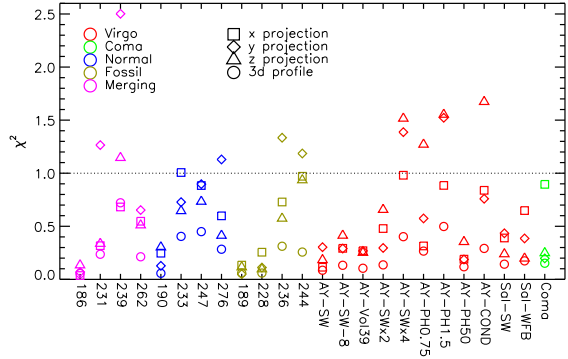}
\caption{The goodness of fit parameter for the different models. It is mostly one group
with extreme merging activity, and the very smooth models for the ``Virgo'' cluster,
with thermal conduction or entropy floor, that the model have problems with fitting.}\label{fig:gof}
\end{center}
\end{figure}

\begin{figure}
\begin{center}
\includegraphics[width=0.45 \textwidth]{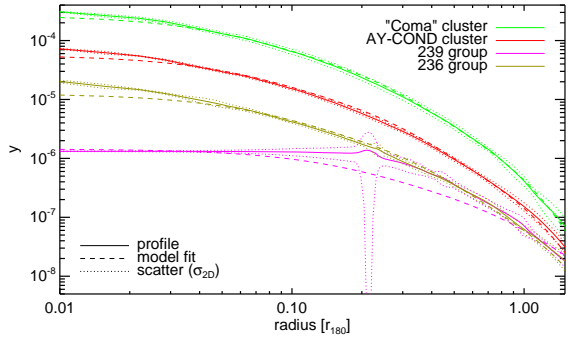}
\caption{Examples of the model fits to the SZ profiles with error bars. Two of the profiles represent
some of the poorest fits, and illustrate why they are poor fits: The 239 group has a poor fit because of
the merging activity, the flatness inside-- and spike at $0.2\,\rhe$ is due to the two central galaxies, separated by approximately $0.4\,\rhe$, while the AY-Cond ``Virgo'' cluster has a poor fit
because the thermal conductivity in the model gives a very smooth cluster with small error bars.
The two others are a representative group and cluster, both with well fitted profiles.}\label{fig:profiles}
\end{center}
\end{figure}

We have found that using a triple exponential model gives an excellent fit to the data:
\begin{align}\nonumber
y^m_{3D}(r) = & Y_m \left[ f_1 \exp(-r/r_0) + f_2 \exp(-r/(\alpha_2 r_0)) \right. \\
                  & \quad \left. + f_3 \exp(-r/(\alpha_3 r_0)) \right]\,,
\end{align}
where $f_1+f_2=1$. While this model at first sight might appear complicated, it has
two attractive features: (i) The 2D profile can be derived in a closed form from
the 3D profile. (ii) Most of the parameters can be fixed to global values, leaving only two
free parameters for fitting all systems considered in this paper. With only
two parameters in the model it can readily be applied to future observations, even if they
only map the SZ profile with a few observational points.
By integrating along one of the axes we find the 2D profile to be
\begin{align}\nonumber
y^m_{2D}(r) = & 2\, Y_m \left[ f_1 K_1(r/r_0) + f_2 K_1(r/(\alpha_2 r_0)) \right. \\ \label{eq:profile}
                  & \qquad \left. + f_3 K_1(r/(\alpha_3 r_0)) \right] r\,,
\end{align}
where $K_n(x)$ is the modified Bessel function of the second kind.
Minimising $\chi^2_{2D}$ simultaneously for all models over the range $0.01\,\rhe< r <1.9\,\rhe$,
while varying $f_i$ and $\alpha_i$ as global parameters, and $Y_m$,
$r_0$ for each model, treating the $x-$, $y-$, and $z-$projections as different
clusters, we find the global best fit parameters to be
\begin{eqnarray}
  f_1 = 0.043 \\
  f_2 =1-f_1 = 0.957 \\
  f_3 = 5.9 \\
  \alpha_2 = 0.36 \\
  \alpha_3 = 0.115 \,,
\end{eqnarray}
In \fig{fig:gof} we see how these yield very reasonable $\chi^2$ for all projections, and
the 3D profiles. The only systems that have a $\chi^2$ significantly larger 
than one are either merging, or characterised by a very smooth profile with almost no
variation in the radial bin, and hence a very small $\sigma(r)$ (see e.g.~\fig{fig:profiles}).

\begin{figure}
\begin{center}
\includegraphics[width=0.45 \textwidth]{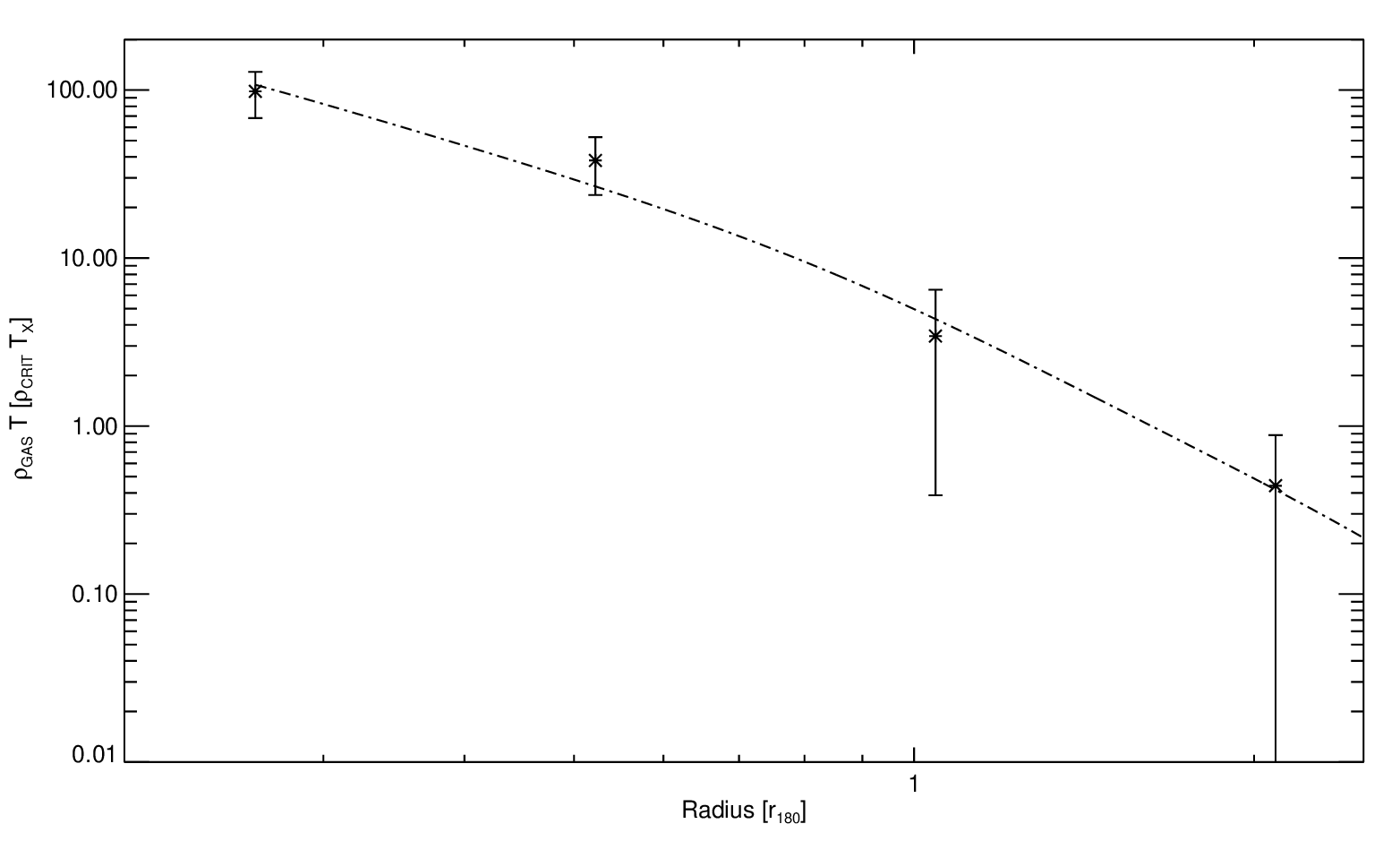}\\
\includegraphics[width=0.45 \textwidth]{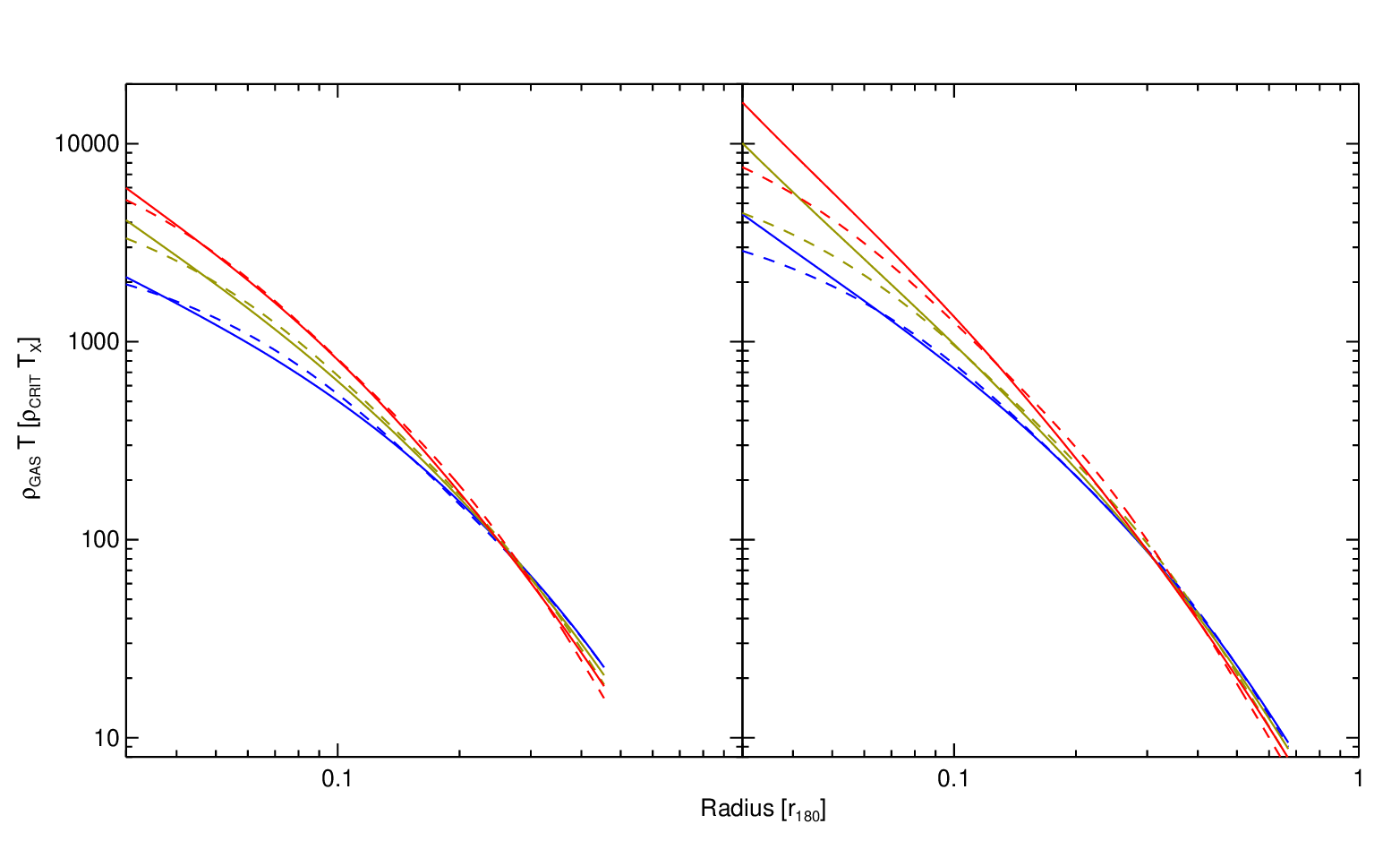}
\caption{The fitted profiles for the observed average profile for massive clusters (top), and for
the constructed profiles assuming hydrostatic equilibrium, and using average dark matter
profiles from the simulations, and observed temperature profiles for groups (bottom left)
and massive clusters (bottom right). The red, yellow, and blue curves are for the average
``virgo'' cluster, fossil group, and normal group DM potentials.}\label{fig:fits}
\end{center}
\end{figure}

\begin{figure}
\begin{center}
\includegraphics[width=0.45 \textwidth]{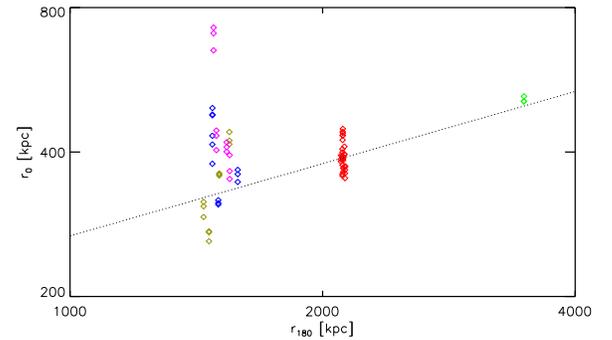}\\
\includegraphics[width=0.45 \textwidth]{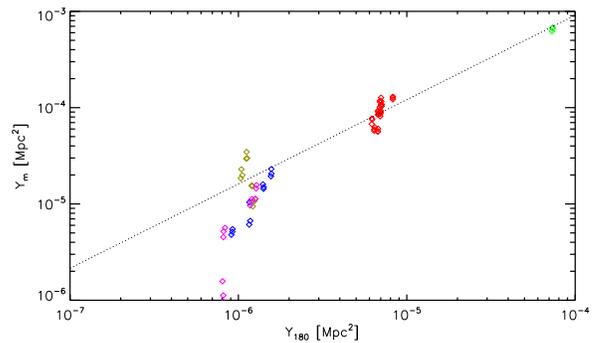}
\caption{The slope $r_0$ (top) and the overall normalisation (bottom) in the model.
The fossil groups act as an extension of the clusters, with the same scaling relation.}\label{fig:propto}
\end{center}
\end{figure}
The two free parameters, the overall scale $r_0$, and the overall normalisation $Y_m$
scale with $\rhe$ and with $Y_{180}$, the integrated SZ effect inside $\rhe$, respectively,
but there is a large scatter between different clusters, where the tightest relation is found for the
fossil groups, that act as an extension of the clusters with almost the same scaling relation.
They are related as
\begin{align}
  Y_m = & 15 \, \left(\frac{Y_{180}}{10^{-6}\, \textrm{Mpc}^2}\right)^{0.9}
                                                         10^{-6}\,\textrm{Mpc}^2\\
  r_0  = & 280 \left(\frac{\rhe}{1000 \textrm{kpc}}\right)^{1/2}\!\! \textrm{kpc}
\end{align}

To push the boundaries of our fitting formula for other systems we have also applied it to
the \citet{Afshordi:2006} data, and the profiles for the centre of the systems, derived from
observed temperature profiles (see \fig{fig:fits}). The average SZ profile is described well
by the formula , while there are problems with fitting the core ($r<0.05\,\rhe$)
of the cluster set of central profiles (the lower right panel in \fig{fig:fits}). Even though some of these
combinations (e.g.~a normal group DM potential combined with a massive cluster temperature profile)
are extreme, the lack of agreement may be because in the central parts of the observed systems
there are significant non-thermal contributions to the pressure balance from e.g.~an AGN
\citep{Roychowdhury:2005} or from cosmic rays \citep{pfrommer:2006}, and hence the
assumption of hydrostatic equilibrium does not apply. Or it may be that the simulated systems
do not include an adequate description of the physical processes in the centre of the systems.
Nonetheless, the contribution of the central $0.05\,\rhe$ to the total SZ signal is
approximately 5\%, and the parameters are not much affected even if we cannot reconstruct the innermost part of the profiles with perfection.

\subsection{Estimating the total mass in a system}
\begin{figure}
\begin{center}
\includegraphics[width=0.45 \textwidth]{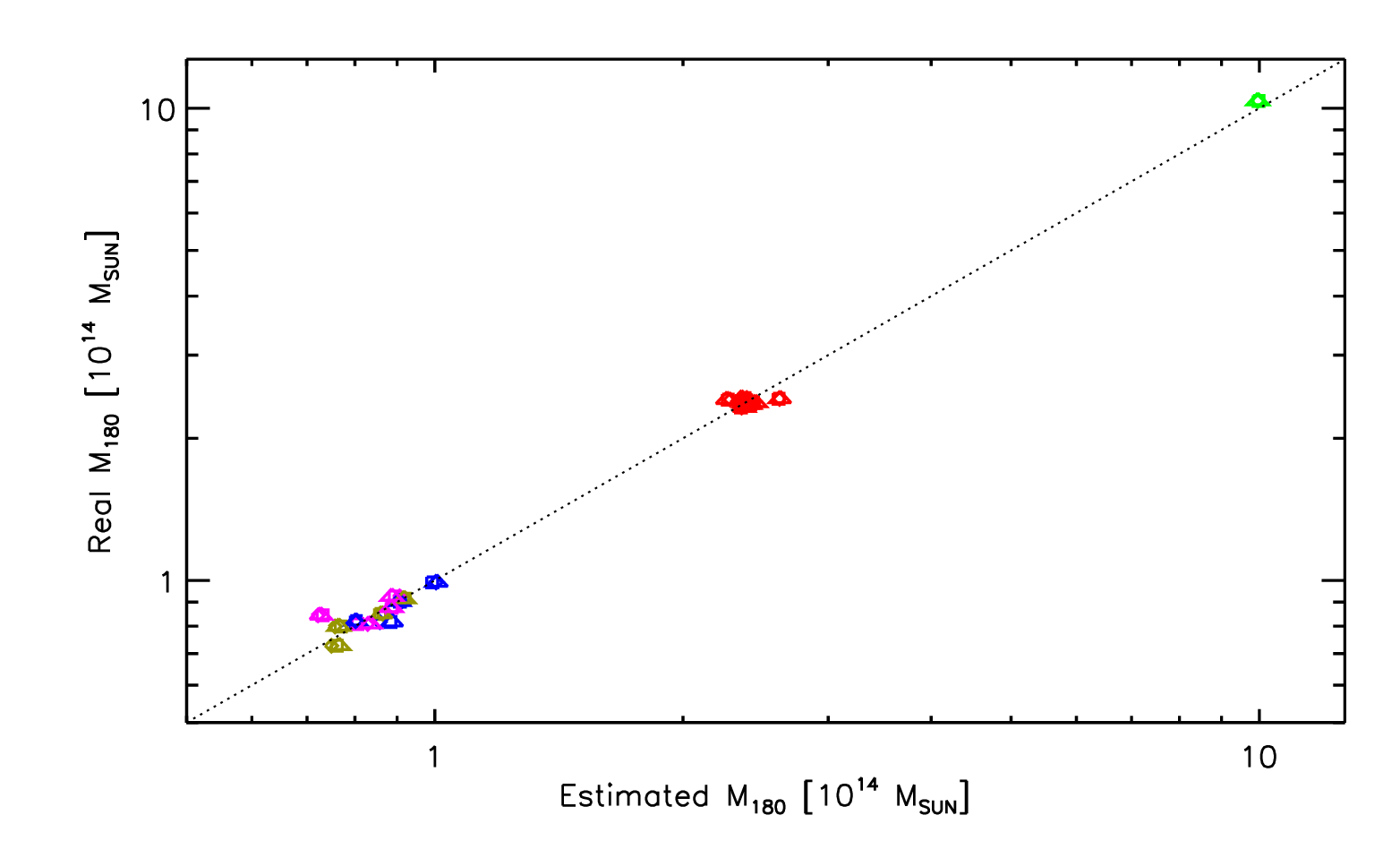}\\
\includegraphics[width=0.45 \textwidth]{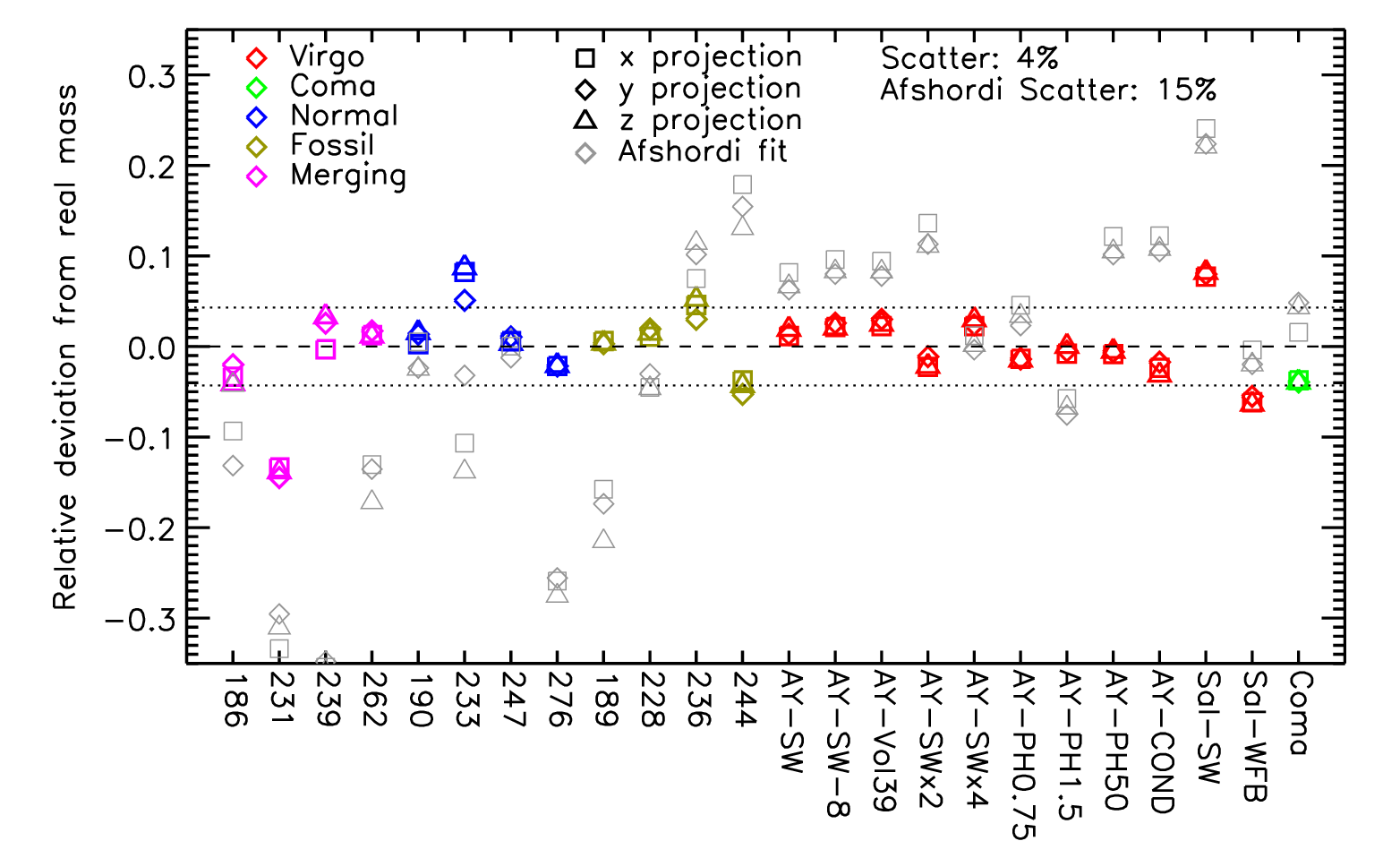}
\caption{The estimated mass (from Eq.~\ref{eq:massestimate}) against the real mass of the
systems (top), and the relative error (bottom). The dotted line indicate the rms scatter. For comparison
we have included the error on the estimated masses using the method of \citet{Afshordi:2007} as
light grey symbols.}\label{fig:massrel}
\end{center}
\end{figure}

The SZ profile of the systems seems to be universally well described by only two parameters,
except for possibly in the central parts of the clusters. It measures the distribution of thermal
energy in the system, and is therefore related to the gravitational potential, if we assume the
system is relaxed. Using assumptions about hydrostatic equilibrium and using a
phenomenological approach several ``Fundamental plane'' relations have been constructed
\citep[e.g.][]{Afshordi:2007,Verde:2002}. The main ingredients for the Fundamental plane
has been the integrated SZ effect and a characteristic scale in the system, for example
$R_{SZ,2}$, the radius enclosing half of the integrated signal.
This has given a relation between the integrated SZ effect, a characteristic scale,
and the total mass with a rough scatter of 14\%
\citep{Afshordi:2007}. With our fit to the profiles we get a very precise measurement
of this characteristic scale, that take into account the different bending of the profiles.
Fitting the total mass as a function of $Y_{180}$ and $r_0$ we find (see
\fig{fig:massrel})\footnote{We have also tried to use $Y_{M}$ instead of $Y_{180}$, but
it does not yield as good a fit.}
\begin{equation}\label{eq:massestimate}
M_{180} = 1.02\,\times 10^{14} M_\odot \, \left(\frac{Y_{180}}{10^{-6} Mpc^2}\right)^{0.57}
                                                     \!\!\left(\frac{r_0}{1000 kpc}\right)^{0.27}\!\!\!\!\!\!\!\!\!\!\!
\end{equation}
with only a 4\% scatter. In \citet{Afshordi:2007} a phenomenological model is used to construct
a set of, essentially, one dimensional models for the clusters. It includes normal or tophat distributed
controlling parameters, with broadening in agreement with observed clusters. To check if the smaller
scatter we see for Eq.~\ref{eq:massestimate} is due to more similarity in the simulated systems, we
have tried to use the fundamental plane of \citet{Afshordi:2007}, on the simulated systems. We find
good agreement with a 15\% scatter for our data (see \fig{fig:massrel}).

\begin{figure}
\begin{center}
\includegraphics[width=0.45 \textwidth]{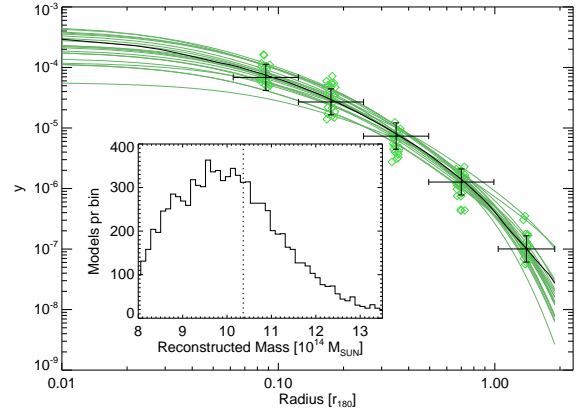}
\caption{A set of mock observations based on the ``Coma'' cluster. The green diamonds,
artificially scattered in the radial direction for visibility, are the mock observations,
while the thin green lines are corresponding fitted profiles. The black line is the
``Coma'' SZ profile, with 1-$\sigma$ errors and the width of each ring overplotted. In the
inset histogram is shown the distribution of the reconstructed masses. The dotted line
is the real mass, which is slightly offset compared to the average reconstructed
mass.}\label{fig:synth}
\end{center}
\end{figure}

The first resolved SZ profiles will be obtained for nearby massive clusters.
To make a crude test for how our fitting formula works on real data, we have
made a set of mock observations taking as a starting point the most massive
cluster in our sample, the ``Coma'' cluster. We use five logaritmically spaced
rings at a distance of $[0.09,0.18,0.35,0.7,1.4]\,\rhe$. The resolution of the
first generation of SZ instruments is limited, and therefore we have chosen
not to include observational points at smaller radii\footnote{Including a central
bin will for this cluster only decrease the error bars though.}.
To get a good measure of the scatter we generated $10^4$ mock observations,
with normal distributed logarithmic error bars of 0.22 dex on the total in
each ring $y_i$, giving an relative error of $\sim50\%$ on $y_i$ itself. For simplicity
we have disregarded any correlation there may exist between the different radial
bins, and artificially have fixed the relative errors to a uniform level.
The five rings are then fitted (see \fig{fig:synth}) using
Eq.~(\ref{eq:profile}) giving $r_0$. The integrated SZ effect $Y_{180}$ for each 
mock observation is found by integrating the fitted profile. Inserting $r_0$ and $Y_{180}$
derived from each mock observation into Eq.~(\ref{eq:massestimate}) we find a reconstructed
mass of $M_{180}=10.0\pm1.4\times10^{14} M_\odot$, or an $14\%$ error on the reconstructed
mass. There is a $4\%$ systematic offset compared to the real mass of $10.4\times10^{14} M_\odot$.
With five rings the error on the total signal is $\sim \sqrt{5}\,\,50\%=22\%$. The
mass goes roughly as $Y_{180}^{0.57}$, and we would expect roughly a $13\%$ error on the mass,
in agreement with what is found. 

\section{Discussion}
In this paper we have constructed a simple empirical model for the radial profile of the
Sunyaev Zeldovich effect in groups and clusters of galaxies. The model has been motivated
by, and validated against a mixture of simulations and observations, and is characterised
by only two parameters: The overall normalisation, and a typical length scale related to the
slope of the profile. It gives a very good fit to the simulated systems, and there is a tight
relation between the parameters and the total mass of the system. Furthermore, the results are
robust to the detailed gas physics employed in the simulations. This can be seen by considering
the subset of the systems, the ``Virgo'' clusters, that are started from the same initial conditions,
but simulated with different implementations of the gas physics (see Table 1). Because of the
different gas physics there is an appreciable scatter in $Y_{180}$, but still the mass ($M_{180}$)
of the clusters is well reconstructed (see \fig{fig:massrel}).

The simulations are in good agreement with an observed average profile extracted from WMAP
data for the outer part of the SZ profiles $0.2\,\rhe<r<2\,\rhe$.
Currently there are no published high resolution observations of the core SZ profile, but we can get
an indirect measure by using temperature profiles extracted from clusters observed in X-rays together
with dark matter potentials from the simulations.
Reconstructing the SZ profile, under the assumption of hydrostatic equillibrium, we see that these
``core profiles'' are relative peaked, compared to the simulations. This is traced to differences in the
temperature profile of the simulated systems, compared to what is observed using X-rays. We have to
await future observations of the resolved cluster cores, to determine to what extent this cusp in the
central part is real, or a result of the hypothesis of hydrostatic equilibrium, which is known to be violated
in the centre of clusters of galaxies, where non-thermal processes such as AGN heating
\citep{Roychowdhury:2005}, and cosmic rays \citep{pfrommer:2006} can play an important role for the
pressure balance.

We stress that the model we have presented here is readily applicable to future observations. This
will give a good proxy for the mass of the observed system. It will also help in reconstructing
the full SZ profile from observations with low spatial resolution, to be used in conjunction with
X-ray observations in the study of cluster dynamics. A fitting routine written in {\tt IDL} for
the profile that can be applied to observed or simulated data  can be found at\\
\url{http://www.phys.au.dk/~haugboel/software.shtml}.

\section*{Acknowledgements}
We thank the Danish Center for Scientific Computing for granting the
computer resources that made this work possible. 
The Dark Cosmology Centre is funded by the Danish National Research Foundation.
This research was supported by the DFG cluster of excellence ``Origin and
structure of the Universe''. KP acknowledges support from the Instrument center for
Danish Astrophysics.

\end{document}